\documentclass[aps, prr, reprint, superscriptaddress, longbibliography]{revtex4-1}
\usepackage{amssymb}
\usepackage{eurosym}
\usepackage{graphicx}
\usepackage{epstopdf}
\usepackage{dcolumn}
\usepackage{bm}
\usepackage{here}
\usepackage[utf8]{inputenc}
\usepackage[T1]{fontenc}
\usepackage{mathptmx}
\usepackage{hyperref}

\begin{document}

\title{Phase transition at 350~K in the Ti$_3$C$_2$T$_x$ MXene: possible sliding (moir\'e) ferroelectricity}
\author{Francesco Cordero}
\email{francesco.cordero@ism.cnr.it}
\affiliation{Istituto di Struttura della Materia-CNR (ISM-CNR), Area della Ricerca di
Roma - Tor Vergata, Via del Fosso del Cavaliere 100, I-00133 Roma, Italy}
\author{Hanna Pazniak}
\affiliation{Universit\'e Grenoble Alpes, CNRS, Grenoble INP, LMGP, 3 parvis Luois N\'eel, F-38000 Grenoble, France}
\author{Thierry Ouisse}
\affiliation{Universit\'e Grenoble Alpes, CNRS, Grenoble INP, LMGP, 3 parvis Luois N\'eel, F-38000 Grenoble, France}
\author{Jesus Gonzalez-Julian}
\affiliation{Department of Ceramics, Institute of Mineral Engineering, RWTH Aachen University, D-52074 Aachen, Germany}
\author{Aldo Di Carlo}
\affiliation{Istituto di Struttura della Materia-CNR (ISM-CNR), Area della Ricerca di
Roma - Tor Vergata, Via del Fosso del Cavaliere 100, I-00133 Roma, Italy}
\author{Viktor Soprunyuk}
\affiliation{Faculty of Physics, University of Vienna, Boltzmanngasse 5, 1090 Wien,
Austria}
\author{Wilfried Schranz}
\affiliation{Faculty of Physics, University of Vienna, Boltzmanngasse 5, 1090 Wien,
Austria}
\date{\today }

\begin{abstract}
A phase transition is found in Ti$_3$C$_2$T$_x$ MXene at 350~K, by measuring the complex Young's modulus of self-standing thick films. A step-like softening and increase of the mechanical losses is found below 350~K, indicative of a phase transition, where the square of the order parameter is coupled to strain. It is argued that it should be a ferroelectric transition, most likely of the sliding (moir\'e) type, due to charge transfer between facing flakes sliding with respect to each other. If the transition will be confirmed to be ferroelectric, Ti$_3$C$_2$T$_x$ will be added to the class of metallic ferroelectrics and open new perspectives of applications, in addition to the numerous already studied.
\end{abstract}

\maketitle


Up to recent times it was believed that ferroelectric transitions cannot
occur in materials with sizes below a certain limit. Indeed, the critical
thickness for a film to sustain ferroelectricity (FE) is found to be of the
order of tens of nanometers in ferroelectrics with perovskite or fluorite
structure \cite{PKS23}. After the prediction of FE with giant
piezoelectricity in the monolayer chalcogenides SnSe, SnS, GeSe, and GeS
\cite{FKY16} and the experimental verifications that CuInP$_{2}$S$_{6}$ has
a FE transition with $T_{\mathrm{C}}$~= 320~K \cite{BHD15b,LYS16b}, the
number of layered van der Waals (vdW) ferroelectrics has increased and
includes $\alpha $-In$_{2}$Se$_{3}$ \cite{WLL18}, $\gamma $-InSe \cite{SJZ23}%
, SnS \cite{HKL20}, SnSe, SnTe \cite{CLL16}, 1T$^{\prime }$-MoTe$_{2}$ \cite%
{YLC19}, 1T$^{\prime }$-MoS$_{2}$ \cite{LCG22}, and 1T$^{\prime }$-WTe$_{2}$
\cite{FZP18,SXS19}. It has also been ascertained that in BN \cite{SWC21}, MoS%
$_{2}$ \cite{WCE22} WTe$_{2}$ and few other systems a new mechanism, other
than lattice instability, is responsible for FE, namely the relative sliding
of the vdW stacked layers, called sliding or moir\'{e} FE \cite%
{YWL18,WL21,LWW23,WGH23} and can even be observed in multilayers of
monoatomic graphene \cite{YDG23}. The phenomenon was first predicted for BN
and mixed BN/graphene bilayers and other vdW bilayers \cite{LW17} and is
based on the charge transfer between the facing atoms in the two layers,
which induces an out-of-plane polarization. The interest in these
ferroelectric mono and bi-layers is great, in view of possible applications
in nanoelectronics \cite{PG22,LWW23}, for example more miniaturized FeRAMs
integrable with Si or other materials thanks to the vdW coupling. Especially
the sliding or Moir\'{e} FE offers unique advantages of robust out-of-plane
polarization \cite{WL21}.

MXenes are other 2D layered materials with chemical formula M$_{n+1}$C$_{n}$T%
$_{x}$ (M standing for a transition metal, X standing for C or N, and T$_{x}$
is surface functional groups). In the MXene structure, the surface groups T$%
_{x}$ (usually --F, =O, --OH), inherited from the etching environment, are
covalently bound to the transition metal and weakly bound via van der
Waals bonds to terminations of neighboring sheets. Such structures are being
extensively studied for a multitude of possible applications in important
sectors \cite{BAB21}, such as electrodes in Li/Na batteries \cite{LLW23},
intermediate layers of perovskite solar cells \cite{APP19}, flexible and
nano-electronics \cite{HG18}, soft robotics \cite{WGT23}, catalysis and many
others \cite{BAB21,LLW23}. However, no ferroelectricity has been found up to
now in MXenes, but only theoretically predicted in Sc$_{2}$CO$%
_{2}$ \cite{CMS17}. On the other hand, Ti$_{3}$C$_{2}$T$_{x}$ monolayers have been
demonstrated to be piezoelectric \cite{TJS21,TSC22}. Piezoelectricity,
namely an induced strain proportional to an applied electric field or
polarization induced by stress, is a consequence of FE, but is also possible
in non-ferroelectrics lacking spontaneous polarization, if their lattice
belongs to certain non-centrosymmetric classes. This is the case of
bi-atomic hexagonal layers and of Ti$_{3}$C$_{2}$T$_{x}$ \cite{TJS21}.

It must be said that the theoretical predictions of 2D FE are more numerous
than the experimental verifications, which are rather challenging. In fact,
these types of FE are probed on mono-, bi- or tri-layer nanoflakes with
techniques such as Piezo-Force Microscopy (PFM) or conductive AFM.

We will show that these types of FE can be probed also with the simple
macroscopic method of measuring the elastic anomaly associated with the FE
transition on thick films of the layered material. We present the first
measurements of the complex Young's modulus versus temperature of
self-standing thick films of the Ti$_{3}$C$_{2}$T$_{x}$ MXene, which exhibit
a clear and robust nearly second order phase transition at $\simeq 350$~K,
and it is argued that, in spite of the good electrical conductivity, the
transition should be ferroelectric, likely of the sliding type.



Ti$_{3}$C$_{2}$T$_{x}$ MXenes were synthesized by selective chemical etching
of the Al layer from a parent Ti$_{3}$AlC$_{2}$ powder precursor using the
minimally intensive layer delamination approach \cite{LAL16}. To produce a
thick Ti$_{3}$C$_{2}$T$_{x}$ film, a suspension of well-delaminated flakes
was vacuum filtered through a nitrocellulose membrane with an average pore
size of 0.22~%
$\mu$%
m. After drying, deposited layers were detached from the membrane, and a
rigid thick free-standing MXene film was obtained. The film thickness was
controlled by the concentration and the amount of filtered suspension.


The complex Young's modulus $E=$ $E^{\prime }+iE^{\prime \prime }$ was
measured in an apparatus where a bar-shaped sample is suspended on thin
thermocouple wires in high vacuum ($<10^{-5}$~mbar), and electrostatically
excited on its free flexural resonant modes \cite{CDC09}. The resonance
frequencies are $f\propto \sqrt{E}$  \cite{NB72}, so that the temperature
dependence of $E$ is evaluated as $E\left( T\right) /E_{0}=$ $\left[ f\left(
T\right) /f_{0}\right] ^{2}$. The elastic energy loss coefficient, $Q^{-1}=$
$E^{\prime \prime }/E^{\prime }$, is measured from the decay of the free
oscillations or the width of the resonance peak. The measurements are made
well within the linear elastic limit. Details on some of the experimental
difficulties encountered and their solutions can be found in the
Supplemental Material (SM). The measurements reported here were made on a
strip 25.5~mm long, 3~mm wide and $\left( 73\pm 4\right) $~$\mu $m thick,
with the first three free flexural resonances clearly visible. In order to
consolidate the film avoiding warping and bubbles (see SM), the sample was
put between two alumina slabs and heated at 1.5~${}^{\circ }$C/min
up to 700~${}^{\circ }$C in $10^{-6}$~mbar, kept 1~h and furnace cooled. The
film became flat and more brittle. Thermal annealing has been reported to
improve electrical conductivity of MXene films and we observed the same
trend in the change in Young's modulus: after annealing it increased from
11.9~GPa to 44~GPa, presumably due to the loss of intercalated species.


The anelastic spectra were measured also using a Perkin-Elmer Diamond DMA in
the three point bending mode, where flexural vibrations are forced at fixed
frequencies in the range 0.1--10~Hz in Ar atmosphere (see SM). The sample
were also characterized before and after the anelastic measurements by XRD,
and Raman spectroscopy (see SM).



\begin{figure}[t]
\begin{center}
\includegraphics[width=8.5cm]{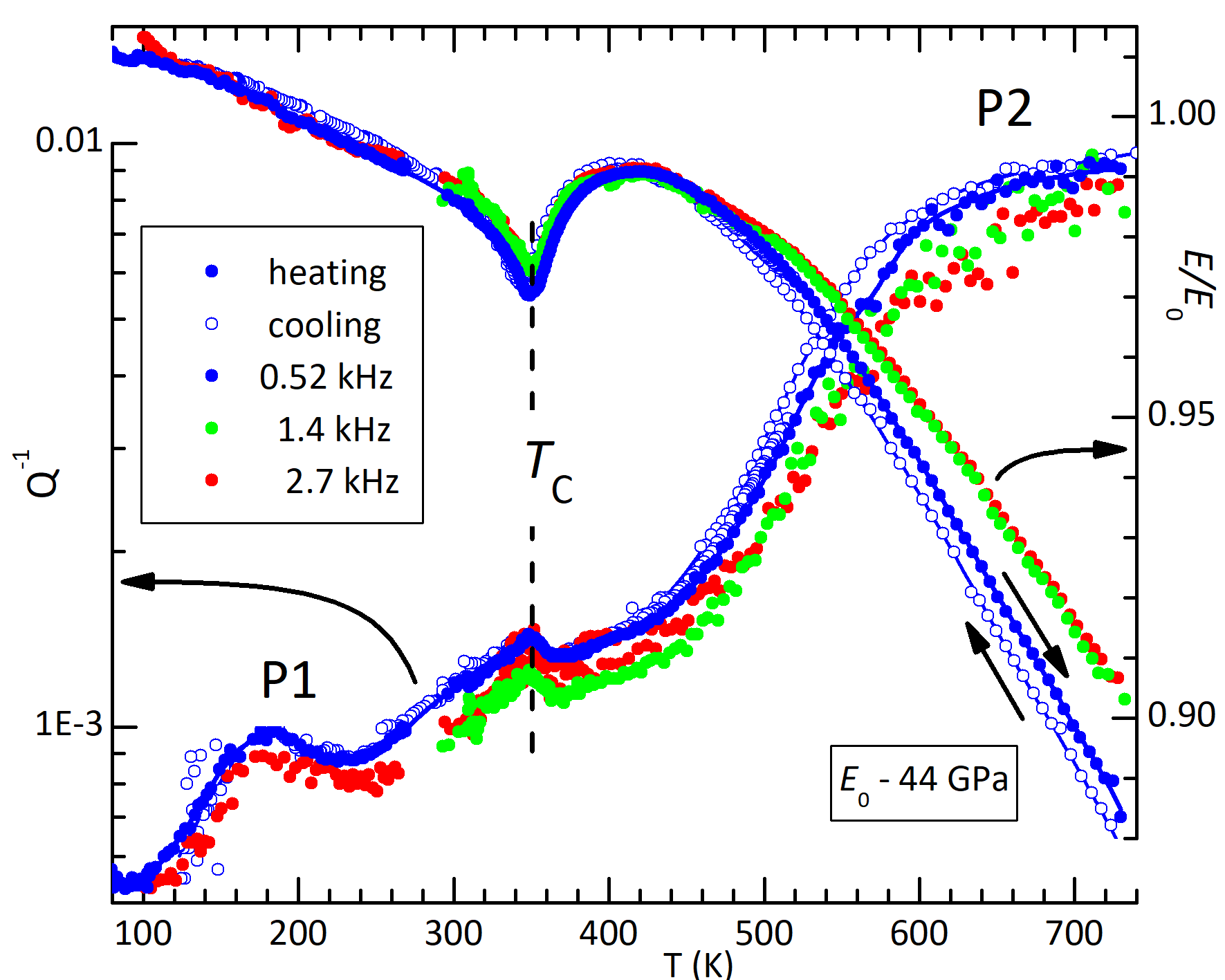}
\end{center}
\caption{Normalized Young's modulus and elastic energy loss measured during
the 4th heating (filled symbols), exciting the 1st, 2nd and 3rd flexural
modes. The empty symbols are the 1st and 3rd coolings.}
\label{fig3fhc}
\end{figure}

Figure \ref{fig3fhc} presents the Young's modulus $E$ and elastic energy
loss $Q^{-1}$ measured during the 1st and 3rd cooling runs (empty symbols),
and subsequent heating (filled symbols). In the last case, in addition to
the fundamental also the 2nd and 3rd flexural modes were excited at higher
frequencies. Until the sample was maintained in vacuum, the measurements
were well repeatable, with some hysteresis between heating and cooling above
room temperature. The main result is the nearly frequency independent
negative step of $E$ and positive step of $Q^{-1}$ below $T_{\mathrm{C}%
}\simeq$ 350~K. The step in $E$ is preceded by a precursor softening
extending tens of kelvin above $T_{\mathrm{C}}$. This is the typical
signature of a phase transition whose squared order parameter is linearly
coupled with strain \cite{Reh73,CS98}.

Above $T_{\mathrm{C}}$ the Young's modulus exhibits some frequency
dispersion, that may be at least partly accounted for by the intense
relaxation process labeled P2. In fact, this broad peak shifts to higher
temperature with increase of frequency, indicating thermal activation of the
spectrum of characteristic times and must be accompanied by a decrease of
the real part, according to the Debye formula $\Delta E\propto \left(
1+i\omega \tau \right) ^{-1}$ in the case of a single relaxation time $\tau $
($\omega =2\pi f$). There is also a minor peak in $Q^{-1}$, labeled P1, too
small to produce visible effects on $E$.

These measurements have been repeated several times, to check their
reproducibility also after exposure to air or immersion in water, as
described in the SM. In brief, it was found that the anelastic spectrum was
reproduced even when measured in 700~mbar\ water vapor up to 430~K, with
little or no intercalation of H$_{2}$O. One day immersion in water resulted
in intercalation of $\sim 0.3$ H$_{2}$O per formula unit, softening the
modulus and strongly depressing the elastic anomaly at $T_{\mathrm{C}}$.
Outdiffusion of water in vacuum was fast above 430~K. The cycle of
experiments was concluded with heating up to 890~K, after which the original
$Q^{-1}$ curve was completely recovered, but $E$ remained $\sim 4\%$ softer
and the step amplitude was smaller than originally (curve 12 in Figs. \ref%
{figDMAres} and S1). This was probably also due to surface oxidation of the
film, revealed by Raman spectroscopy after completing the anelastic
measurements.

\begin{figure}[t]
\begin{center}
\includegraphics[width=8.5cm]{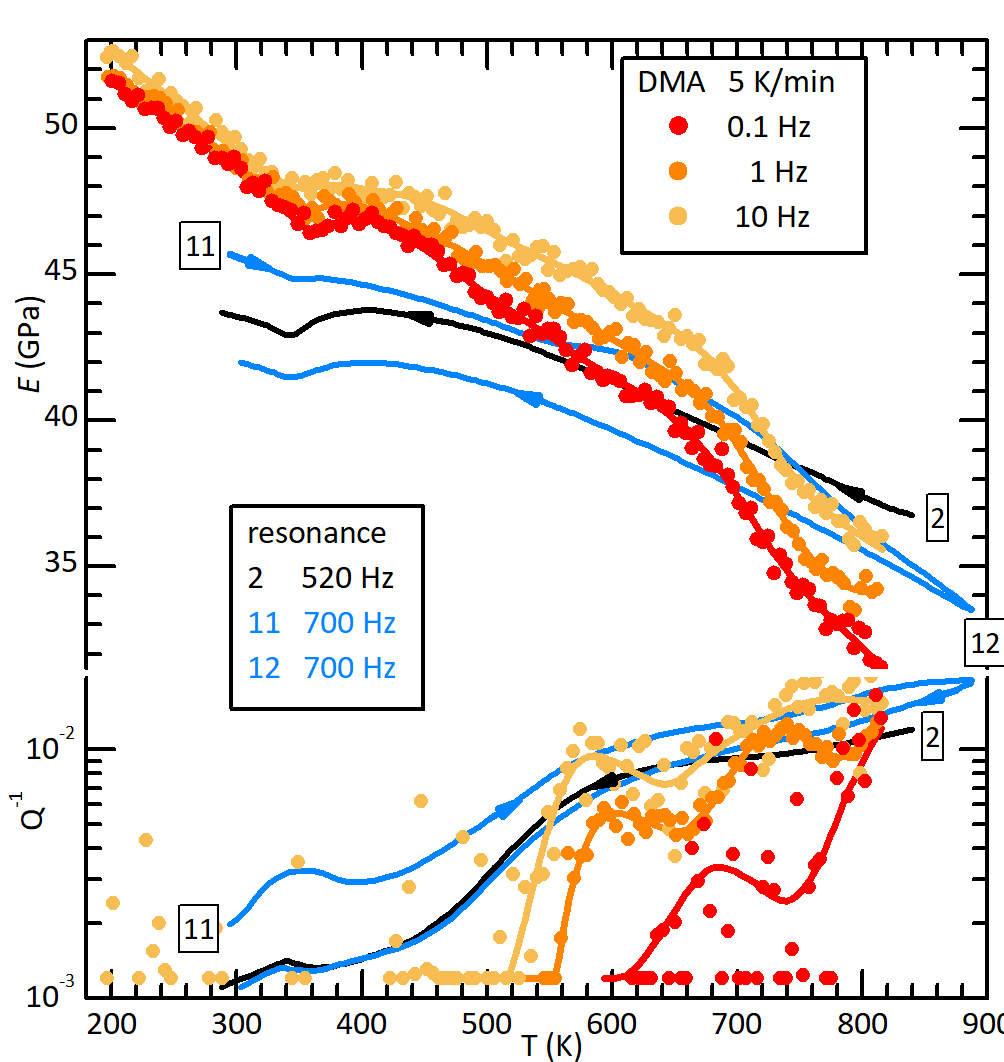}
\end{center}
\caption{Young's modulus and elastic energy loss measured with DMA at three
frequencies heating at 5~K/min. The previous first and last runs with the
resonance method are also shown, numbering the temperature scans in the same
manner as in Fig. S1.}
\label{figDMAres}
\end{figure}

The DMA measurements substantially confirm all the previous ones with the
resonance method, and provide additional information on the relatively slow
dynamics characterizing the transition.
A strip of as-prepared material had at room temperature a Young's modulus as
low as 4~GPa, which increased up to 25~GPa after warming up to 495~K (not
shown), confirming the consolidation effect of heating the as-deposited
MXene sheet in vacuum or inert atmosphere. Figure \ref{figDMAres}\ presents
a DMA run from 200 to 810~K at 5~K/min and 0.1, 1, 10~Hz on the same sample
that had been measured with the resonance method. Two months passed between
the two sets of measurements. For comparison, the first and last runs with
the resonance method are also shown, with the same labeling of Fig. S1. The
major result is the confirmation of the steplike softening below 350~K. The
magnitude of the modulus measured with the two methods is also in fair
agreement, considering that the $E\left( T\right) $ curves are quite
reproducible until the sample is measured and continuously kept in HV, but
otherwise they can vary considerably.

The DMA measurements put in evidence a progressive stiffening of $E$ with
increase of frequency above $T_{\mathrm{C}}$ and a broad step around 700~K,
that had appeared also in the last resonance heating (curve 11).

\begin{figure}[t]
\begin{center}
\includegraphics[width=8.5cm]{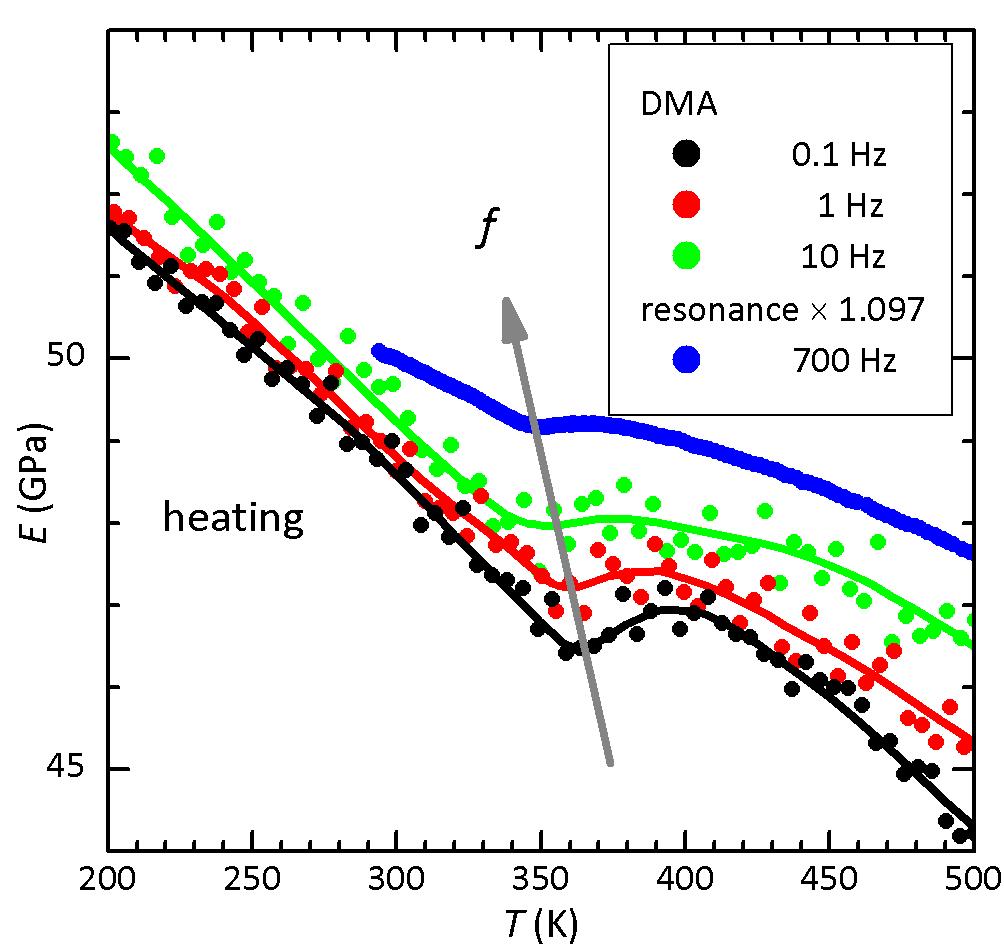}
\end{center}
\caption{Frequency dependence of the step in the Young's modulus at the
phase transition. The resonance data are multiplied by 1.097.}
\label{fig-f-LK}
\end{figure}

More interesting is the frequency dispersion at the phase transition, put in
evidence in Fig. \ref{fig-f-LK}. The last heating with the resonance method
(curve 11 of Fig. \ref{figDMAres}) is multiplied by 1.097 for clarity. A
10\% difference in the magnitude of the Young's moduli measured with the two
methods may be due to imperfect shape and homogeneity of the sample and is
well within the 20\% estimated absolute accuracy of the DMA method \cite%
{SKT12}. In addition, it may be partly due to the variability of the modulus
following exposure to air and high temperature cycles.


XRD and Raman spectra were measured again after the above experiments,
as described in the SM. The
XRD spectra did not show any new peak, excluding decomposition and formation
of new phases within the bulk of the film, but were shifted, indicating a
decrease of the $c$ lattice parameter from 26.35 to 20.22~\AA\ . This is due
to the loss of the initial intercalants. The Raman spectra, probing only the
surface, exhibited new peaks attributable to TiO$_{2}$ (atanase), absent in
the XRD spectrum, indicating that surface oxidation occurred.


The major result of these anelastic measurements is the presence of a
previously unnoticed nearly second
order phase transition at $T_{\mathrm{C}}\simeq 350$~K.

The elastic anomaly consists of a negative step of $\sim 4\%$ in the Young's
modulus $E\ $ below $T_{\mathrm{C}}$, accompanied by a positive step in the
elastic energy loss $Q^{-1}$ and preceded by precursor softening extending
tens of kelvin above $T_{\mathrm{C}}$. This type of anomaly is
characteristic of a phase transition whose order parameter (OP) $P$ is
coupled with strain $\epsilon $ through a term $\propto \epsilon P^{2}$ in
the free energy \cite{Reh73}. This is the case of a ferroelectric transition
where $P$ is the electric polarization, but also of a (anti)ferromagnetic or
antiferrodistortive transition (like octahedral tilting in perovskites \cite%
{KSS00}).

A magnetic transition can be excluded, since MXenes are mainly Pauli
paramagnets. In some cases the magnetization of Ti$_{3}$C$_{2}$T$_{x}$
is very small down to 60~K, where a small kink may indicate a
paramagnetic-to-antiferromagnetic transition \cite{JWM22}.

Regarding the possibility of an antiferrodistortive transition in vdW solids,
we know of only two possibilities in layered organic-inorganic halide perovskites.
One is coordinated Jahn-Teller distortions of the BX$_6$ octahedra \cite{SJK20},
but this is not possible with Ti$^4+$. The other is tilting of the octahedra
\cite{NL21}. This type of
transition is common in perovskites AMX$_{3}$ \cite{Gla72}, with short
strong M-X bonds and longer weaker A-X bonds. When the network of rigid
corner-sharing octahedra MX$_{6}$, sharing only their vertices X, cannot
follow the thermal contraction of the more anharmonic A-X bonds, the rigid
octahedra tilt in order to fit into the smaller lattice, giving rise to the
antiferrodistortive transition. In Ti$_{3}$C$_{2}$T$_{x}$ there are only
Ti-C bonds, and the network of edge sharing CTi$_{6}$ octahedra \cite{HLH17}
has neither the structural flexibility nor a driving force for an
antiferrodistortive instability. Indeed, deviations from the in-plane
hexagonal structure have never been reported, neither with traditional X-ray
diffraction nor with its pair distribution function analysis \cite{SBN14}.

Therefore, the remaining possibility is that the transition is
ferroelectric. Indeed, the elastic anomaly in Ti$_{3}$C$_{2}$T$_{x}$ is very
similar to that found in classical ferroelectrics, for example in BaTiO$%
_{3-\delta }$, observable both in the insulating state with $\delta =0$ \cite%
{Cor18} and when the material is made metallic by introducing O vacancies
\cite{CTC19}. The similarity is not limited to the steps in elastic modulus
and damping, but includes a precursor softening extending at least tens of
kelvin above $T_{\mathrm{C}}$ \cite{Cor18,CTS23}. The major difference
between the elastic anomalies in Ti$_{3}$C$_{2}$T$_{x}$ and BaTiO$_{3-\delta
}$ is the magnitude of the softening, which is up to $50\%$ in BaTiO$_{3}$
and $\sim 4\%$ in Ti$_{3}$C$_{2}$T$_{x}$.

The ferroelectric nature of the transition, however, cannot be ascertained
with the usual electric methods, since Ti$_{3}$C$_{2}$T$_{x}$ is a good
electric conductor and any applied electric field is shielded by the free
charge carriers. There are only few known metallic ferroelectrics, among
which is the vdW layered WTe$_{2}$ \cite{SXS19}. Probing FE in vdW layered
materials is achieved through sophisticated experiments on mono-, bi- and
tri-layer nanoflakes, but the task is particularly challenging in the case
of electrically conducting materials. The present results show that the
possible ferroelectric transition in Ti$_{3}$C$_{2}$T$_{x}$ can be probed
also with a macroscopic measurement like the temperature dependence of the
complex elastic modulus of a self-standing thick film. Not only it is not
necessary to probe the polarization switching at the atomic level, but the
good electric conductivity of the sample, which hinders the transition to
the usual electrical methods, does not affect the elastic properties, which
therefore reveal the elastic anomaly from the coupling between strain and
polarization.

As far as the elastic anomaly is concerned, FE in Ti$_{3}$C$_{2}$T$_{x}$ may
be due to the usual lowering of the symmetry of the paraelectric phase, like
Ti off-centering in BaTiO$_{3}$ or vertical shifting of the intermediate Se
plane in the vdW In$_{2}$Se$_{3}$ \cite{XZW18}. In the present case, a
possibility would be vertical shifting of the C planes with respect to the
Ti planes. This type of polar structure, however, has never been reported
for Ti$_{3}$C$_{2}$T$_{x}$; rather, it has been found that monolayers are
piezoelectric with null polarization in the absence of strain \cite{TJS21}.
This corresponds to intrinsic piezoelectricity rather than switchable
spontaneous polarization within a monolayer.

Ti$_{3}$C$_{2}$T$_{x}$ presents similarities with bilayer WTe$_{2}$, which
is both semi-metallic and ferroelectric with $T_{\mathrm{C}}\simeq 350$~K
\cite{FZP18,YWL18}. The origin of its FE has been identified with the
relative sliding between layers, which forms out-of-plane electric dipoles,
due to the transfer of electronic charge between the facing atoms. The fact
that the electric dipoles are formed between pairs of layers and
approximately perpendicular to them explains why the polarization is not
completely screened by the in-plane conductivity \cite{YWL18}. Relative
sliding of two facing monolayers can switch the polarization over an
extended area without requiring atomic displacements within the monolayers.
A clear indication that the transition is due to an interlayer mechanism,
and not intrinsic of each layer, is the fact that it is not observed in the
as-prepared state, abundantly intercalated, and disappears by intercalating
water.

The Curie temperature in sliding FE is well above room temperature, even
though the sliding barrier is of $\lesssim $~meV \cite{YWL18}. This is very
convenient for applications, since it combines low switching fields with
robust ferroelectricity, and has been explained in the framework of
continuum electromechanics as a consequence of the large in-plane rigidity
of the layers \cite{TB23}. For Ti$_{3}$C$_{2}$T$_{x}$ there would be the
complication that the facing species are not the ordered planes of surface
Ti atoms but the --OH, =O and --F terminations. We are not aware of reports
of sliding FE between non uniformly terminated layers, but will discuss the
elastic anomaly in this light, though the analysis will be valid for any
type of FE. In fact, we can exploit the fact that in semiconducting bilayer
WSe$_{2}$, exhibiting sliding FE, the spontaneous polarization $P\left(
T\right) $ can be fitted with the usual Landau free energy with terms up to
the 6th power of $P$ \cite{LLL22b},
\begin{equation}
F=\frac{a}{2}\left( T-T_{\mathrm{C}}\right) P^{2}+\frac{B}{4}P^{4}+\frac{C}{6%
}P^{6},  \label{F}
\end{equation}%
generally used for describing any first ($B<0$) or second ($B>0$) order
ferroelectric transition. Once the transition is phenomenologically
described in terms of Eq. (\ref{F}), the elastic anomaly can be evaluated
introducing a suitable coupling between stress/strain and order parameter $P$%
. We disregard a possible\ coupling of $P$ with the "intrinsic"
piezoelectricity found in the non-centrosymmetric monolayer of Ti$_{3}$C$%
_{2} $T$_{x}$ \cite{TJS21}, with in-plane polarization $p$, because $p$ and $%
P$ are approximately perpendicular to each other and physically separated.
In fact, $p$ is in-plane and confined by the good electrical conductivity
within the layers, while $P$ is approximately perpendicular to and between
the layers. In addition, while $p$ is due to the atomic displacements within
the layers, $P$ is considered to arise from interlayer electronic transfer
with minimal relative atomic displacements within layers \cite{LWW23}.

In order to evaluate the effect of the transition on the elastic modulus, we
include in the free energy the elastic and the mixed terms containing both $P
$ and strain $\varepsilon $ or, considering the Gibbs free energy $%
G=F-\sigma \varepsilon $, the terms containing $P$ and stress $\sigma $.
Without entering in the details of the mechanism causing the polarization,
we can safely include the electrostrictive coupling $Q\sigma P^{2}$, always
allowed \cite{LG77,Reh73}, while the piezoelectric coupling $d\sigma p$ is
with the intrinsic $p$, so that
\begin{eqnarray}
G &=&\frac{a}{2}\left( T-T_{\mathrm{C}}\right) P^{2}+\frac{B}{4}P^{4}+\frac{C%
}{6}P^{6}-\frac{S^{0}}{2}\sigma ^{2}-Q\sigma P^{2}+ \\
&&+\frac{1}{2\chi _{p}}p^{2}-d\sigma p,  \label{G}
\end{eqnarray}%
where the susceptibility $\chi _{p}$ of $p$, nearly independent of $T$, is
included.

The elastic compliance $S=1/E$ is calculated as usual \cite{Reh73,SL98,CCT16}
as $S=$ $d\varepsilon /d\sigma $ with $\varepsilon =$ $-\partial G/\partial
\sigma $ and one obtains (see SM)
\begin{equation}
\begin{array}{c}
S\left( T>T_{\mathrm{C}}\right) =S^{0}+\chi _{p}d^{2} \\
S\left( T<T_{\mathrm{C}}\right) =S^{0}+\chi _{p}d^{2}+\Delta s \\
\Delta S=\frac{2Q^{2}}{B\sqrt{\frac{T_{\mathrm{C}}+\Delta T-T}{\Delta T}}}%
,~~\Delta T=\frac{B^{2}}{4aC}%
\end{array}
\label{ds-2nd-lin2}
\end{equation}%
The constant term $\chi _{p}d^{2}$ is the softening due to the intrinsic
piezoelectricity, and just reorganizes the background compliance $S^{0}$.
The coupling to $P$ does not produce any effect in the paraelectric phase ($%
T>T_{\mathrm{C}}$ for $B<0$ and up to $T_{\mathrm{C}}+\Delta T$ for $B<0$),
where $P$ is null apart from fluctuations, and produces a steplike softening
$\Delta S$ in the ferroelectric phase. The step is constant of magnitude $%
2Q^{2}/B$ if $C=0$ and acquires a cusped shape for $C>0$. This type of
elastic anomaly is the same for any OP that is coupled quadratically to
stress/strain, including magnetization and the rotation angle in
antiferrodistortive transition, as in SrTiO$_{3}$ \cite{KSS00}. The
softening is due to the relaxation of the order parameter upon application
of stress, with consequent additional strain. In the FE case, the softening
is of piezoelectric origin \cite{CCT16}.

In view of the marked dependence on frequency shown in Fig. \ref{fig-f-LK},
the time dependence of the response of $P$ to stress should be taken into
account. This is usually done in the framework of the Landau-Khalatnikov
theory, assuming that $\dot{P}=-\left( P-P_{0}\right) /\left( \chi L\right) $%
, where $L$ is a nearly temperature independent kinetic coefficient and $%
\chi $ the susceptibility $\sim \left\vert T-T_{\mathrm{C}}\right\vert ^{-1}$%
, yielding a relaxation time $\tau =\tau _{0}/\left( T_{\mathrm{C}}-T\right)
$. Introducing this time dependence of $P$ into the elastic response to a
periodic excitation with angular frequency $\omega $ yields a complex
compliance $\Delta S/\left( 1+i\omega \tau \right) $ \cite{KSS00,YVK06b} and
therefore a frequency dependence of the real part%
\begin{eqnarray}
\Delta S\left( \omega \right) &=&\Delta S/\left[ 1+\left( \omega \tau
\right) ^{2}\right]  \label{DS(w)} \\
\tau &=&\tau _{0}/\left( T_{\mathrm{C}}-T\right)  \nonumber
\end{eqnarray}

\begin{figure}[t]
\begin{center}
\includegraphics[width=8.5cm]{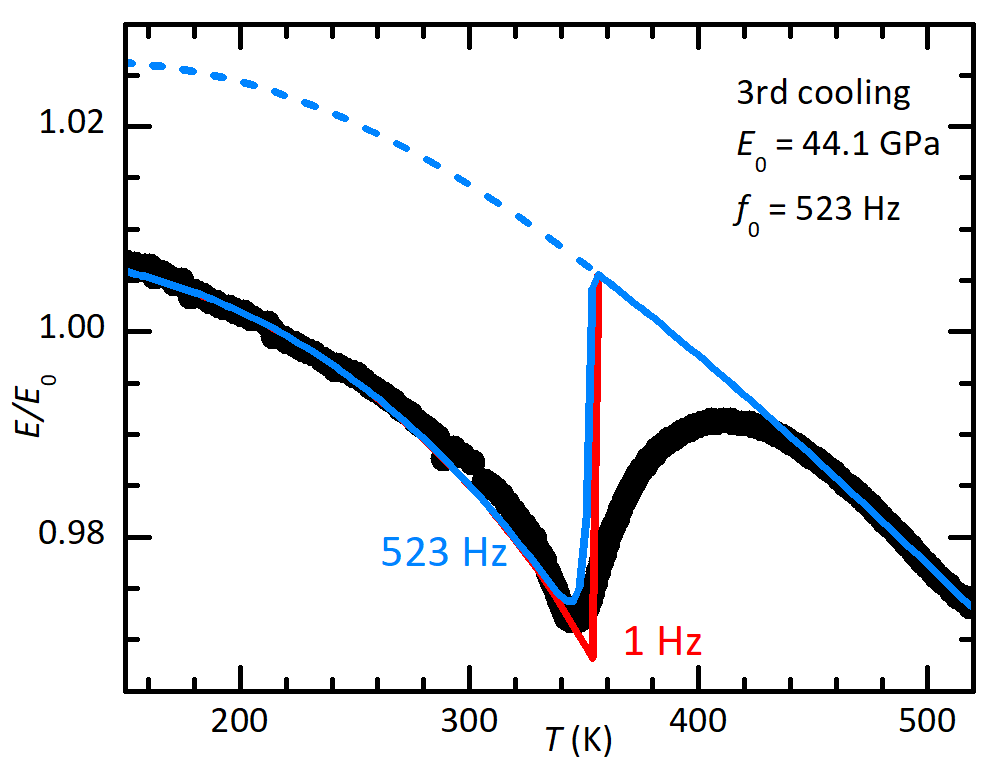}
\end{center}
\caption{Fit of the elastic anomaly presented in Fig. \protect\ref{fig3fhc},
measured at 523~Hz during the 3rd cooling, using Eq. (\protect\ref%
{ds-2nd-lin2}) with the parameters $\Theta =$ 490~K; $T_{\mathrm{C}}=$
354~K, $\Delta T=$ 70~K; $\Delta s\times E_{0}=$ 0.039, $\protect\tau _{0}=$
$7\times 10^{-4}$~s. The red line is calculated at 1~Hz, The dashed line is
the background $E^{0}$.}
\label{figfitbk}
\end{figure}

In order to fit the anomaly of the Young's modulus $E$ with an expression
like Eqs. (\ref{ds-2nd-lin2}--\ref{DS(w)}), it is necessary to establish the
background variation of $E\left( T\right) $. This usually is a linear
stiffening during cooling, due to anharmonic effects, with quantum
saturation below some temperature $\Theta $, and can be written in terms of
the compliance as \cite{SWT91}
\begin{equation}
1/E^{0}=S^{0}=S_{0}+S_{1}/\tanh \left( \Theta /T\right) ~.  \label{Sbg}
\end{equation}

However, additional phenomena affect $E\left( T\right) $, as apparent in
Figs. \ref{fig3fhc} and \ref{figDMAres}, and therefore a fit cannot be
extended above 500~K, where a smooth frequency dependent kink is evident in
Fig. \ref{fig3fhc}. The low frequency experiment, spanning two decades in $f$%
, actually shows dependence of the background $E^{0}$ down to room
temperature and is also noisier than in the resonance experiments. For this
reason, we present a fit of the elastic anomaly measured during the 3rd
cooling run in resonance, when the sample had not yet been subjected to the
various treatments with water. The best fit with Eqs. (\ref{ds-2nd-lin2}--%
\ref{Sbg}) is found with the parameters indicated in the caption to Fig. \ref%
{figfitbk} (blue line). The red line is calculated at 1~Hz, in order to show
the effect of the relaxation of the order parameter:\ increasing frequency,
the minimum shifts to temperatures lower than $T_{\mathrm{C}}$ and its
magnitude decreases. This is of the same type as observed in Fig. \ref%
{fig-f-LK}, though its full magnitude cannot be reproduced with Eq. (\ref%
{DS(w)}). It is possible that a contribution to softening below the
transition is also given by the motion of domain walls, which is typically
larger at lower frequency \cite{KSS00}, and may well explain the larger
magnitude of the step at lower frequency.

As anticipated, Eq. (\ref{ds-2nd-lin2}) does not contain the contribution of
fluctuations above $T_{\mathrm{C}}$, and therefore cannot reproduce the
evident precursor softening. Precursor softening is found commonly and in
the classical ferroelectric BaTiO$_{3}$ it extends even to hundreds of
kelvin above $T_{\mathrm{C}}$ \cite{CTS23}. It is notable that the Curie
temperatures reported up to now for sliding ferroelectrics, WSe$_{2}$ \cite%
{LLL22b} and WTe$_{2}$\ \cite{FZP18,YWL18}, are $T_{\mathrm{C}}\simeq 350$~K
as in our case.

These measurements demonstrate that nanoscale phenomena like FE in mono and
bi-layers of vdW solids, generally studied with advanced techniques at the
nanoscale, can be probed with mechanical spectroscopy on thick films. Notice
that it is not necessary that the direction of the interlayer polarization
is coherent along different pairs of layers, since strain is coupled to $%
P^{2}$ and strains from opposite polarizations do not cancel with each other.

The dissipation $Q^{-1}\left( T\right) $ of Ti$_{3}$C$_{2}$T$_{x}$ presents
a positive step below $T_{\mathrm{C}}$, which is the normal behaviour from
the interaction between stress and the walls between the domains formed in
the low-temperature phase \cite{CS98,KSS00,Cor18}. With increase of
temperature, the dissipation is dominated by intense relaxation processes
with a broad spectrum of characteristic times toward high temperature (P2 in
Fig. \ref{fig3fhc}). Among the possible contributions to these losses are
relative sliding of the flakes, hopping of the residual intercalated
species, diffusion of the terminating molecules, and motion of point and
extended defects within the monolayers.

The magnitude of the Young's modulus of the film consolidated in high vacuum
at 970~K, $E=$ $40-50$~GPa, is larger than that achieved in Ti$_{3}$C$_{2}$T$%
_{x}$\ densified by addition of sodium carboxymethyl cellulose and sodium
borate (28~GPa) \cite{WLC21b} or carboxymethylated cellulose nanofibrils
(42~GPa) \cite{TVR19}, but the films are fragile and $E$\ is still one order
of magnitude smaller than the 330~GPa found with AFM on single flakes \cite%
{LLA18}. The large discrepancy between the in-plane Young's modulus of
single and aggregated flakes must be due to their easy relative sliding.


We measured for the first time the temperature dependence of the complex
Young's modulus $E$ of the Ti$_{3}$C$_{2}$T$_{x}$ MXene in the form of
self-standing thick films. In the as-prepared state the films have a very
low $E\sim 4-10$~GPa, that may be increased up to $40-50$~GPa by annealing
in high vacuum at 970~K, so reducing the amount of intercalated species.
This is still lower than 330~GPa of a single flake, presumably due to an
easy relative sliding of the flakes, but allows reproducible anelastic
spectra to be measured up to 900~K. The major feature is a phase transition
below $T_{\mathrm{C}}\sim 350$~K, producing a steplike softening of $\sim
4\% $ of $E$ and increase of dissipation, which is robust against thermal
treatments up to 900~K in vacuum or inert gas. The nature of the phase
transition is discussed and it is concluded that it should be ferroelectric,
based on the fact that the available structural and magnetic measurements
exclude deviations from the hexagonal structure and magnetism at room
temperature. Particular attention has been devoted to the mechanism of
sliding ferroelectricity, where sliding of the layers with respect to each
other induces charge transfer between atoms/molecules belonging to the
facing layers and hence to interlayer polarization. Notably, $T_{\mathrm{C}%
}\simeq 350$~K is the same of other two known sliding ferroelectrics, WSe$%
_{2}$ has \cite{LLL22b} and WTe$_{2}$ \cite{FZP18,YWL18}. This would add a
new feature and perspectives of applications to the already numerous ones of
Ti$_{3}$C$_{2}$T$_{x}$ MXene. In this view, the simple fact that this
transition is slightly above room temperature calls for further
investigations of its nature.

FC and ADC acknowledge the precious technical assistance of M.P. Latino (CNR-ISM).


\begin{thebibliography}{10}

\bibitem{PKS23}
Sung~Hyuk Park, Jae~Young Kim, Jae~Yong Song, and Ho~Won Jang.
\newblock {Overcoming Size Effects in Ferroelectric Thin Films}.
\newblock {\em Adv. Phys. Res.}, 2:2200096, 2023.

\bibitem{FKY16}
Ruixiang Fei, Wei Kang, and Li~Yang.
\newblock {Ferroelectricity and Phase Transitions in Monolayer Group-IV
  Monochalcogenides}.
\newblock {\em Phys. Rev. Lett.}, 117:097601, 2016.

\bibitem{BHD15b}
A.~Belianinov, Q.~He, A.~Dziaugys, P.~Maksymovych, E.~Eliseev, A.~Borisevich,
  A.~Morozovska, J.~Banys, Y.~Vysochanskii, and S.~V. Kalinin.
\newblock {CuInP$_2$S$_6$ Room Temperature Layered Ferroelectric}.
\newblock {\em Nano Lett.}, 15:3808, 2015.

\bibitem{LYS16b}
Fucai Liu, Lu~You, Kyle~L. Seyler, Xiaobao Li, Peng Yu, Junhao Lin, Xuewen
  Wang, Jiadong Zhou, Hong Wang, Haiyong He, Sokrates~T. Pantelides, Wu~Zhou,
  Pradeep Sharma, Xiaodong Xu, Pulickel~M. Ajayan, Junling Wang, and Zheng Liu.
\newblock {Room-temperature ferroelectricity in CuInP$_2$S$_6$ ultrathin
  flakes}.
\newblock {\em Nat. Commun.}, 7:12357, 2016.

\bibitem{WLL18}
Siyuan Wan, Yue Li, Wei Li, Xiaoyu Mao, Wenguang Zhu, and Hualing Zeng.
\newblock {Room-temperature ferroelectricity and a switchable diode effect in
  two-dimensional $\alpha$-In$_2$Se$_3$ thin layers}.
\newblock {\em Nanoscale}, 10:14885, 2018.

\bibitem{SJZ23}
Fengrui Sui, Min Jin, Yuanyuan Zhang, Ruijuan Qi, Yu-Ning Wu, Rong Huang,
  Fangyu Yue, and Junhao Chu.
\newblock {Sliding ferroelectricity in van der Waals layered $\gamma$-InSe
  semiconductor}.
\newblock {\em Nat. Commun.}, page~36, 2023.

\bibitem{HKL20}
Naoki Higashitarumizu, Hayami Kawamoto, Chien-Ju Lee, Bo-Han Lin, Fu-Hsien Chu,
  Itsuki Yonemori, Tomonori Nishimura, Katsunori Wakabayashi, Wen-Hao Chang,
  and Kosuke Nagashio.
\newblock {Purely in-plane ferroelectricity in monolayer SnS at room
  temperature}.
\newblock {\em Nat. Commun.}, 11:2428, 2020.

\bibitem{CLL16}
Kai Chang, Junwei Liu, Haicheng Lin, Na~Wang, Kun Zhao, Anmin Zhang, Feng Jin,
  Yong Zhong, Xiaopeng Hu, Wenhui Duan, Qingming Zhang, Liang Fu, Qi-Kun Xue,
  Xi~Chen, , and Shuai-Hua Ji.
\newblock {Discovery of robust in-plane ferroelectricity in atomic-thick SnTe}.
\newblock {\em Science}, 353:274, 2016.

\bibitem{YLC19}
Shuoguo Yuan, Xin Luo, Hung~Lit Chan, Chengcheng Xiao, Yawei Dai, Maohai Xie,
  and Jianhua Hao.
\newblock {Room-temperature ferroelectricity in MoTe$_2$ down to the atomic
  monolayer limit}.
\newblock {\em Nat. Commun.}, 10:1775, 2019.

\bibitem{LCG22}
A.~Lipatov, P.~Chaudhary, Z.~Guan, H.~Lu, G.~Li, O.~Cr{\'e}gut, K.~D. Dorkenoo,
  R.~Proksch, S.~Cherifi-Hertel, D.~F. Shao, E.~Y. Tsymbal, J.~{\'I}{\~n}iguez,
  A.~Sinitskii, and A.~Gruverman.
\newblock {Direct observation of ferroelectricity in two-dimensional MoS$_2$}.
\newblock {\em npj 2D Mater. Appl.}, 6:18, 2022.

\bibitem{FZP18}
Z.~Fei, W.~Zhao, T.~A. Palomaki, B.~Sun, M.~K. Miller, Z.~Zhao, J.~Yan, X.~Xu,
  and D.~H. Cobden.
\newblock {Ferroelectric switching of a two-dimensional metal}.
\newblock {\em Nature}, 560:336, 2018.

\bibitem{SXS19}
Pankaj Sharma, Fei-Xiang Xiang, Ding-Fu Shao, Dawei Zhang, Evgeny~Y. Tsymbal,
  Alex~R. Hamilton, and Jan~Seidel and.
\newblock {A room-temperature ferroelectric semimetal}.
\newblock {\em Sci. Adv.}, 5:eaax5080, 2019.

\bibitem{SWC21}
M.~Vizner Stern, Y.~Waschitz, W.~Cao, I.~Nevo, K.~Watanabe, T.~Taniguchi,
  E.~Sela, M.~Urbakh, O.~Hod, and M.~Ben Shalom.
\newblock {Interfacial ferroelectricity by van der Waals sliding}.
\newblock {\em Science}, 372:1462, 2021.

\bibitem{WCE22}
Astrid Weston, Eli~G. Castanon, Vladimir Enaldiev, F{\'a}bio Ferreira,
  Shubhadeep Bhattacharjee, Shuigang Xu, H{\'e}ctor Corte-Le{\'o}n, Zefei Wu,
  Nicholas Clark, Alex Summerfield, Teruo Hashimoto, Yunze Gao, Wendong Wang,
  Matthew Hamer, Harriet Read, Laura Fumagalli, Andrey~V. Kretinin, Sarah~J.
  Haigh, Olga Kazakova, A.~K. Geim, Vladimir~I. Falko, and Roman Gorbachev.
\newblock {Interfacial ferroelectricity in marginally twisted 2D
  semiconductors}.
\newblock {\em Nat. Nanotechnol.}, 17:390, 2022.

\bibitem{YWL18}
Qing Yang, Menghao Wu, and Ju~Li.
\newblock {Origin of Two-Dimensional Vertical Ferroelectricity in WTe$_2$
  Bilayer and Multilayer}.
\newblock {\em J. Phys. Chem. Lett.}, 9:7160, 2018.

\bibitem{WL21}
Menghao Wu and Ju~Li.
\newblock {Sliding ferroelectricity in 2D van der Waals materials: Related
  physics and future opportunities}.
\newblock {\em PNAS}, 118:e2115703118, 2021.

\bibitem{LWW23}
Shuhui Li, Feng Wang, Yanrong Wang, Jia Yang, Xinyuan Wang, Xueying Zhan, Jun
  He, and Zhenxing Wang.
\newblock {Van der Waals Ferroelectrics: Theories, Materials and Device
  Applications}.
\newblock {\em Advan. Mater.}, page 2301472, 2023.

\bibitem{WGH23}
Zhe Wang, Zhigang Gui, and Li~Huang.
\newblock {Sliding ferroelectricity in bilayer honeycomb structures: A
  first-principles study}.
\newblock {\em Phys. Rev. B}, 107:035426, 2023.

\bibitem{YDG23}
Liu Yang, Shiping Ding, Jinhua Gao, and Menghao Wu.
\newblock {Atypical Sliding and Moir\'e Ferroelectricity in Pure Multilayer
  Graphene}.
\newblock {\em Phys. Rev. Lett.}, 131:096801, 2023.

\bibitem{LW17}
Lei Li and Menghao Wu.
\newblock {Binary Compound Bilayer and Multilayer with Vertical Polarizations:
  Two-Dimensional Ferroelectrics, Multiferroics, and Nanogenerators}.
\newblock {\em ACS Nano}, 11:6382, 2017.

\bibitem{PG22}
Jacob Parker and Yi~Gu.
\newblock {van der Waals ferroelectrics: Progress and an outlook for future
  research directions}.
\newblock {\em J. Appl. Phys.}, 132:160901, 2022.

\bibitem{BAB21}
Anha Bhat, Shoaib Anwer, Kiesar~Sideeq Bhat, M.~Infas~H. Mohideen, Kin Liao,
  and Ahsanulhaq Qurashi.
\newblock {Prospects challenges and stability of 2D MXenes for clean energy
  conversion and storage applications}.
\newblock {\em npj 2D Mater. Appl.}, 5:61, 2021.

\bibitem{LLW23}
Guohao Li, Shuhan Lian, Jie Wang, Guanshun Xie, Nan Zhang, and Xiuqiang Xie.
\newblock {Surface chemistry engineering and the applications of MXenes}.
\newblock {\em J. Materiomics}, 2023.

\bibitem{APP19}
A.~Agresti, A.~Pazniak, S.~Pescetelli, A.~Di Vito, D.~Rossi, A.~Pecchia, M.~Auf
  der Maur, A.~Liedl, R.~Larciprete, Denis~V. Kuznetsov, D.~Saranin, and A.~Di
  Carlo.
\newblock {Titanium-carbide MXenes for work function and interface engineering
  in perovskite solar cells}.
\newblock {\em Nat. Mater.}, 18:1228, 2019.

\bibitem{HG18}
Kanit Hantanasirisakul and Yury Gogotsi.
\newblock {Electronic and Optical Properties of 2D Transition Metal Carbides
  and Nitrides (MXenes)}.
\newblock {\em Adv. Mater.}, 30:1804779, 2018.

\bibitem{WGT23}
Yizhou Wang, Tianchao Guo, Zhengnan Tian, Lin Shi, Sharat~C. Barman, and
  Husam~N. Alshareef.
\newblock {MXenes for soft robotics}.
\newblock {\em Matter}, 6:2807, 2023.

\bibitem{CMS17}
Anand Chandrasekaran, Avanish Mishra, and Abhishek~Kumar Singh.
\newblock {Ferroelectricity, Antiferroelectricity, and Ultrathin 2D
  Electron/Hole Gas in Multifunctional Monolayer MXene}.
\newblock {\em Nano Lett.}, 4:3290, 2017.

\bibitem{TJS21}
Dongchen Tan, Chengming Jiang, Nan Sun, Jijie Huang, Zhe Zhang, Qingxiao Zhang,
  Jingyuan Bu, Sheng Bi, Qinglei Guo, and Jinhui Song.
\newblock {Piezoelectricity in monolayer MXene for nanogenerators and
  piezotronics}.
\newblock {\em Nano Energy}, 90:106528, 2021.

\bibitem{TSC22}
Dongchen Tan, Nan Sun, Long Chen, Jingyuan Bu, and Chengming Jiang.
\newblock {Piezoelectricity in Monolayer and Multilayer Ti$_3$C$_2$T$_x$
  MXenes: Implications for Piezoelectric Devices}.
\newblock {\em ACS Appl. Nano Mater.}, 5:1034, 2022.

\bibitem{LAL16}
A.~Lipatov, M.~Alhabeb, M.~R. Lukatskaya, A.~Boson, Y.~Gogotsi, and
  A.~Sinitskii.
\newblock {Effect of Synthesis on Quality, Electronic Properties and
  Environmental Stability of Individual Monolayer Ti$_3$C$_2$ MXene Flakes}.
\newblock {\em Adv. Electron. Mater.}, 2:1600255, 2016.

\bibitem{CDC09}
F.~Cordero, L.~Dalla Bella, F.~Corvasce, P.~M. Latino, and A.~Morbidini.
\newblock {An insert for anelastic spectroscopy measurements from 80~K to
  1100~K}.
\newblock {\em Meas. Sci. Technol.}, 20:015702, 2009.

\bibitem{NB72}
A.~S. Nowick and B.~S. Berry.
\newblock {\em {Anelastic Relaxation in Crystalline Solids}}.
\newblock Academic Press, New York, 1972.

\bibitem{Reh73}
W.~Rehwald.
\newblock {The Study of Structural Phase Transitions by Means of Ultrasonic
  Experiments}.
\newblock {\em Adv. Phys.}, 22:721, 1973.

\bibitem{CS98}
M.~A. Carpenter and E.~H.~K. Salje.
\newblock {Elastic anomalies in minerals due to structural phase transitions}.
\newblock {\em Eur. J. Mineral.}, 10:693--812, 1998.

\bibitem{SKT12}
W.~Schranz, H.~Kabelka, and A.~Tr{\"o}ster.
\newblock {Superelastic softening of ferroelastic multidomain crystals}.
\newblock {\em Ferroelectrics}, 426:242, 2012.

\bibitem{KSS00}
A.~V. Kityk, W.~Schranz, P.~Sondergeld, D.~Havlik, E.~K.~H. Salje, and J.~F.
  Scott.
\newblock {Low-frequency superelasticity and nonlinear elastic behavior of
  SrTiO$_3$ crystals}.
\newblock {\em Phys. Rev. B}, 61:946, 2000.

\bibitem{JWM22}
Haolin Jin, Kai Wang, Zhongquan Mao, Lingyun Tang, Jiang Zhang, and Xi~Chen.
\newblock {The structural, magnetic, Raman and electrical transport properties
  of Mn intercalated Ti$_3$C$_2$T$_x$}.
\newblock {\em J. Phys.: Condens. Matter}, 34:015701, 2022.

\bibitem{SJK20}
Dipayan Sen, Gour Jana, Nitin Kaushal, Anamitra Mukherjee, and Tanusri
  Saha-Dasgupta.
\newblock {Intrinsic ferromagnetism in atomically thin two-dimensional
  organic-inorganic van der Waals crystals}.
\newblock {\em Phys. Rev. B}, 102:054411, 2020.

\bibitem{NL21}
J.~A. McNulty and P.~Lightfoot.
\newblock {Structural chemistry of layered lead halide perovskites containing
  single octahedral layers}.
\newblock {\em IUCrJ}, 8:485, 2021.

\bibitem{Gla72}
A.~M. Glazer.
\newblock {The classification of tilted octahedra in perovskites}.
\newblock {\em Acta Cryst. B}, 28:3384, 1972.

\bibitem{HLH17}
Tao Hu, Zhaojin Li, Minmin Hu, Jiemin Wang, Qingmiao Hu, Qingzhong Li, and
  Xiaohui Wang.
\newblock {Chemical Origin of Termination-Functionalized MXenes:
  Ti$_3$C$_2$T$_2$ as a Case Study}.
\newblock {\em J. Phys. Chem. C}, 121:19254, 2017.

\bibitem{SBN14}
Chenyang Shi, Majid Beidaghi, Michael Naguib, Olha Mashtalir, Yury Gogotsi, and
  Simon J.~L. Billinge.
\newblock {Structure of Nanocrystalline Ti$_3$C$_2$ MXene Using Atomic Pair
  Distribution Function}.
\newblock {\em Phys. Rev. Lett.}, 112:125501, 2014.

\bibitem{Cor18}
F.~Cordero.
\newblock {Quantitative evaluation of the piezoelectric response of unpoled
  ferroelectric ceramics from elastic and dielectric measurements: Tetragonal
  BaTiO$_3$}.
\newblock {\em J. Appl. Phys.}, 123:094103, 2018.

\bibitem{CTC19}
F.~Cordero, F.~Trequattrini, F.~Craciun, H.~T. Langhammer, D.~A.~B. Quiroga,
  and Jr. P.~S.~Silva.
\newblock {Probing ferroelectricity in highly conducting materials through
  their elastic response: Persistence of ferroelectricity in metallic
  BaTiO$_{3-d}$}.
\newblock {\em Phys. Rev. B}, 99:064106, 2019.

\bibitem{CTS23}
Francesco Cordero, Francesco Trequattrini, Paulo~Sergio da~Silva, Jr., Michel
  Venet, Oktay Aktas, and Ekhard K.~H. Salje.
\newblock {Elastic precursor effects during Ba$_{1-x}$Sr$_x$TiO$_3$
  ferroelastic phase transitions}.
\newblock {\em Phys. Rev. Research}, 5:013121, 2023.

\bibitem{XZW18}
Jun Xiao, Hanyu Zhu, Ying Wang, Wei Feng, Yunxia Hu, Arvind Dasgupta, Yimo Han,
  Yuan Wang, David~A. Muller, Lane~W. Martin, PingAn Hu, and Xiang Zhang.
\newblock {Intrinsic Two-Dimensional Ferroelectricity with Dipole Locking}.
\newblock {\em Phys. Rev. Lett.}, 120:227601, 2018.

\bibitem{TB23}
Ping Tang and Gerrit~E.W. Bauer.
\newblock {Sliding Phase Transition in Ferroelectric van der Waals Bilayers}.
\newblock {\em Phys. Rev. Lett.}, 130:176801, 2023.

\bibitem{LLL22b}
Yang Liu, Song Liu, Baichang Li, Won~Jong Yoo, and James Hone.
\newblock {Identifying the Transition Order in an Artificial Ferroelectric van
  der Waals Heterostructure}.
\newblock {\em Nano Lett.}, 22:1265, 2022.

\bibitem{LG77}
M.~E. Lines and A.~M. Glass.
\newblock {\em {Principles and Applications of Ferroelectrics and Related
  Materials}}.
\newblock Oxford University Press, Oxford, 1977.

\bibitem{SL98}
B.~A. Strukov and A.~P. Levanyuk.
\newblock {\em {Ferroelectric Phenomena in Crystals}}.
\newblock Springer, Heidelberg, 1998.

\bibitem{CCT16}
F.~Cordero, F.~Craciun, F.~Trequattrini, and C.~Galassi.
\newblock {Piezoelectric softening in ferroelectrics: ferroelectric versus
  antiferroelectric PbZr$_{1-x}$Ti$_{x}$O$_{3}$}.
\newblock {\em Phys. Rev. B}, 93:174111, 2016.

\bibitem{YVK06b}
R~M. Yevych, Yu.~M. Vysochanskii, M.~M. Khoma, and S.~I. Perechinskii.
\newblock {Lattice instability at phase transitions near the Lifshitz point in
  proper monoclinic ferroelectrics}.
\newblock {\em J. Phys.: Condens. Matter}, 18:4047--4064, 2006.

\bibitem{SWT91}
E.K.H. Salje, B.~Wruck, and H.~Thomas.
\newblock {Order-parameter saturation and low-temperature extension of Landau
  theory}.
\newblock {\em Z. Phys. B - Condensed Matter}, 82:399, 1991.

\bibitem{WLC21b}
Sijie Wan, Xiang Li, Ying Chen, Nana Liu, Yi~Du, Shixue Dou, Lei Jiang, and
  Qunfeng Cheng.
\newblock {High-strength scalable MXene films through bridging-induced
  densification}.
\newblock {\em Science}, 374:96, 2021.

\bibitem{TVR19}
Weiqian Tian, Armin VahidMohammadi, Michael~S. Reid, Zhen Wang, Liangqi Ouyang,
  Johan Erlandsson, Torbj{\"o}rn Pettersson, Lars W{\aa}gberg, Majid Beidaghi,
  and Mahiar~M. Hamedi.
\newblock {Multifunctional Nanocomposites with High Strength and Capacitance
  Using 2D MXene and 1D Nanocellulose}.
\newblock {\em Adv. Mater.}, 31:1902877, 2019.

\bibitem{LLA18}
A.~Lipatov, H.~Lu, M.~Alhabeb, B.~Anasori, A.~Gruverman, Y.~Gogotsi, and
  A.~Sinitskii.
\newblock {Elastic properties of 2D Ti$_3$C$_2$T$_x$ MXene monolayers and
  bilayers}.
\newblock {\em Sci. Adv.}, 4:eaat0491, 2018.

\end{thebibliography}

\end{document}


\begin{center}
{\Large Supplemental Material to "Phase transition at 350~K in the Ti$_{3}$C$%
_{2}$T$_{x}$ MXene: possible sliding (moir\'{e}) ferroelectricity"}

\medskip

Francesco Cordero, Hanna Pazniak, Thierry Ouisse, Jesus Gonzalez-Julian,
Aldo Di Carlo, Viktor Soprunyuk, Wilfried Schranz

\bigskip
\end{center}

\section{Anelastic measurements - free flexural resonance}

The complex Young's modulus $E=$ $E^{\prime }+iE^{\prime \prime }$ was
measured in an apparatus where a bar-shaped sample is suspended on thin
thermocouple wires with a diameter of 50~$\mu $m in high vacuum ($<10^{-5}$%
~mbar), and electrostatically excited on its free flexural resonant modes.
The vibration is detected inserting the exciting electrode in a resonant
circuit at $\sim 13$~MHz, whose frequency is modulated by the change of the
capacitance between sample and electrode, and then demodulated to provide a
signal proportional to the sample vibration \cite{CDC09}. The resonance
frequency of the fundamental flexural mode is \cite{NB72}
\begin{equation}
f=1.028\frac{h}{l^{2}}\sqrt{\frac{E}{\rho }}~,  \label{EqFlex}
\end{equation}%
where $l$, $h$, $\rho $ are the sample's length, thickness and density. The
next four modes have frequencies in the ratios $1:$ $2.8:$ $5.49:$ $9.07:$
13.5, allowing the flexural modes to be distinguished from other types of
vibrations. This is important with sample like these films, where the
thickness is much less than the other dimensions and a large variety of
modes can be excited. The variation of the sample dimensions due to the
thermal expansion is usually negligible in comparison with respect to the
change of the elastic moduli, so that the temperature dependence of $E$ is
evaluated as $E\left( T\right) /E_{0}=$ $\left[ f\left( T\right) /f_{0}%
\right] ^{2}$.

\section{Preliminary anelastic measurements\protect\bigskip\ and flattening
anneal}

The suspension thermocouple wires have a diameter of 50~$\mu $m, and are in
correspondence with the nodes of the 1st and 5th flexural modes, nearly
coinciding, in order to minimize the coupling with the sample vibration. The
sample must be fixed to the wires with drops of Ag paint at the edge,
otherwise it would be attracted by the electrode. With samples as thin as
the films studied here, having masses two orders of magnitude smaller than
the typical samples used in the same apparatus, the intrinsic dissipation $%
Q^{-1}$ is easily exceeded by extrinsic effects, such as the coupling with
the wires, the Ag drops and the He exchange gas necessary to cool the sample
below room temperature. At the lowest values of dissipation measured on
these samples, $Q^{-1}<$ $10^{-3}$, exceeding 0.02~mbar of He already
resulted in a clear increase of $Q^{-1}$. The $Q^{-1}\left( T\right) $
curves below room temperature have been corrected subtracting the step
measured after the introduction of He. Also the resonance frequencies are
affected by extrinsic effects, resulting in frequent jumps, whose amplitude
is much larger for the vibration modes other than the 1st and 5th, for which
the wires are not at the nodes of the vibration. After a number of
measurements, we found a manner of coping with this problem, at least for
the fundamental resonance, by applying a slight mechanical excitation to the
sample holder before each measurement.

The first attempts to measure the complex Young's modulus of Ti$_{3}$C$_{2}$T%
$_{x}$ strips 10 to 80~$\mu $m thick and 12 to 28\ mm long failed, because
the strips bent considerably but reproducibly while heated or cooled in HV.
In addition, there were too many resonances and it was difficult to single
out those due to the flexural modes. Some of the resonances increased their
frequency after heating beyond 700~K in HV, indicating near doubling of the
elastic modulus after reaching 850~K. During this run, bubbles had developed
on both faces with diameters of 1--2\ mm, probably due to intercalated
species that were prevented from being desorbed by a very small connected
porosity.

The flexural resonance measurements reported here were made on a strip
25.5~mm long, 3~mm wide and $\left( 73\pm 4\right) $~$\mu $m thick, with the
first three free flexural resonances clearly visible. In order to
consolidate the film avoiding warping and bubbles, the sample was put
between two alumina slabs 0.5~thick and heated at 1.5~${}^{\circ }$C/min up
to 700~${}^{\circ }$C in $10^{-6}$~mbar, kept 1~h and furnace cooled. The
film became flat and very brittle, and its Young's modulus increased from
11.9~GPa to 44~GPa. The consolidated sample could be measured exciting the
1st, 2nd, 3rd and 5th free flexural modes, whose frequencies ratios to the
fundamental were within 6.7\%, 4.7\% and 1.1\% of the theoretical values,
respectively. The larger deviations were found in the 2nd and 3rd modes, for
which the suspension wires did not match the nodes.

\section{Effect of water vapor and immersion in water on the anelastic
spectra}

After the anelastic measurements of Fig. 1 two additional runs were
conducted between room temperature and 890 and 410~K, in order to tune the
experiment and verify that 14~h in air did not affect the anelastic
spectrum. Afterwards the focus was to ascertain a possible influence of the
terminations T$_{x}$ and intercalated species on the phase transition.
Figure S1 collects some of the salient temperature runs, including those in
Figs. 1,2 (curves 2--4). The higher initial $Q^{-1}$ of run 1 is attributed
to the uncured fixing drops of Ag paint.

\begin{figure}[t]
\begin{center}
\includegraphics[width=10cm]{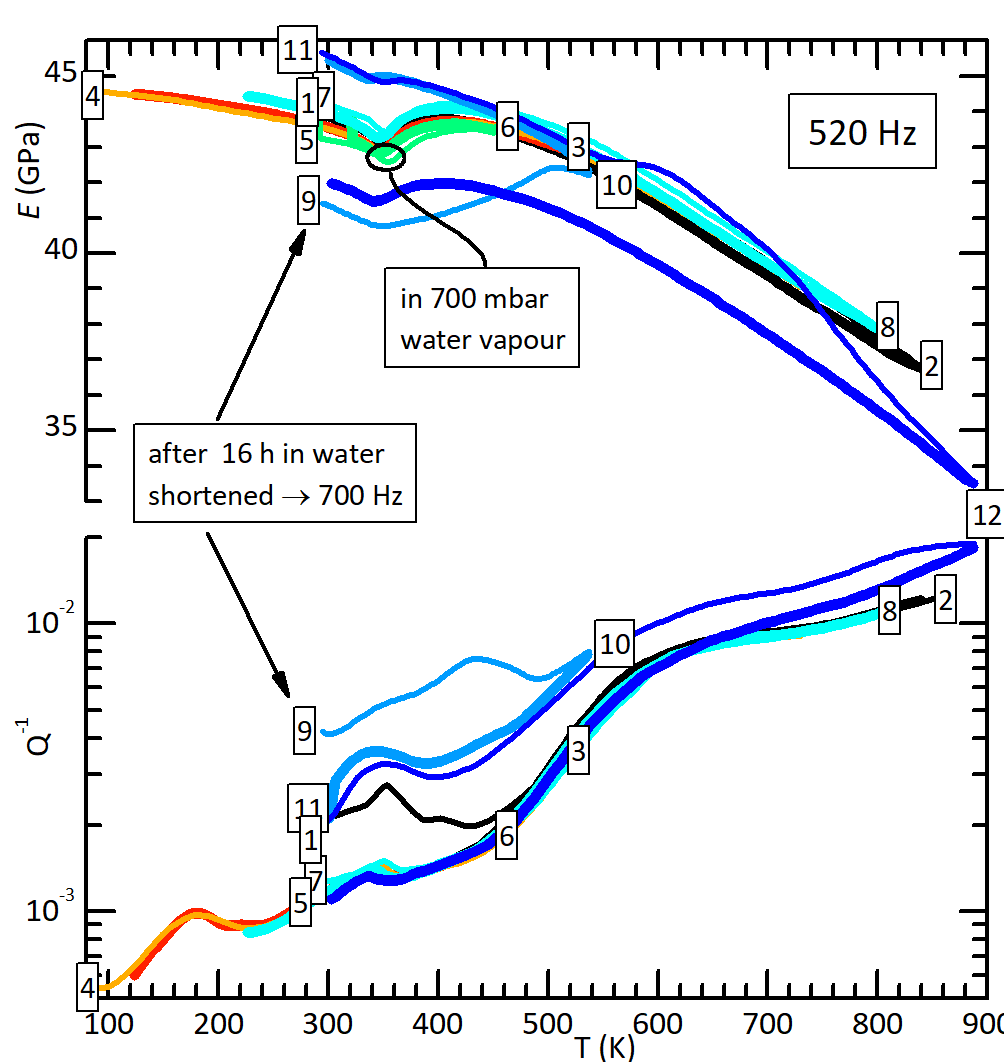}
\end{center}
\caption{Young's modulus and
elastic energy loss measured during various heating (thin lines) and cooling
(thick lines) runs exciting the 1st flexural mode. The numbers are placed at
the start of the runs. After immersion in water, the sample was shortened
and the resonance frequency passed from 520 to 702~Hz.}
\label{figall}
\end{figure}

As a first attempt to intercalate H$_{2}$O, some drops of deionized water
were put on the bottom of the quartz tube enclosing the insert, and the
sample was measured in 700~mbar of air saturated with water vapor up to
450~K, maintained $1/2$~h, and cooled to room temperature, for a total of
4~h in water vapor. Apart from a constant decrease of 1.5\%, due to the
effect of gas pressure on the vibration, the modulus presented exactly the
same anomaly at $T_{\mathrm{C}}$ (curves 5, 6). The dissipation is not shown
because it was large and featureless, due the damping effect of gas.
Immediately after evacuating the humid air, modulus and dissipation were
back at the original values, indicating that the intercalation of water had
been absent or minimal. Repeating the measurements in HV produced the same
curves as before exposure to humidity (curves 7, 8).

The next step was immersing the sample in deionized water for 2~h at 60~${%
{}^{\circ }}$C + 16~h at 22~${{}^{\circ }}$C with an intake $n\simeq $ $0.46$
of intercalated H$_{2}$O per formula unit, estimated from the mass increase.
During 1.5~h in air the sample had lost $\sim 0.053$ H$_{2}$O per formula
unit and an unknown amount in additional 2.5~h during which the sample was
shortened due to difficulties in mounting. The amount of intercalated H$_{2}$%
O at the the start of run 9 was therefore $n\sim $ 0.32 per formula unit.
The effect (curve 9) is to increase the damping and soften the modulus and,
from the recovery of both the $E\left( T\right) $ and $Q^{-1}\left( T\right)
$ curves, it seems that the loss of intercalated H$_{2}$O is accelerated
above 430~K in HV. The heating was limited to 536~K and curve 10 during
cooling was higher than originally both in $E$ and $Q^{-1}$, with a reduced
step in $E$ below $T_{\mathrm{C}}$. During a final cycle up to 890~K in HV
the original $Q^{-1}$ curve was recovered, while $E$ remained $\sim 4\%$
softer and the step amplitude was smaller than originally.

\section{Dynamic Mechanical Analyzer}

The anelastic spectra were measured also using a Perkin-Elmer Diamond DMA.
Attempts at using the tension mode, where the sample is clamped at both
ends, failed because clamping broke the samples. Therefore we used three
point bending, where the sample is suspended on two metallic edges 5~mm
apart and pushed in the centre by an edge-terminated rod. An alternate force
is applied to the rod, in order to induce a vibration with an amplitude of
up to 60~$\mu $m, corresponding to a maximum tensile strain of $7\times
10^{-4}$, plus a constant force of 30~mN. All the measurements were
performed in Ar atmosphere.

\section{XRD, Raman, SEM}

The structural properties of the free-standing films were studied by X-ray
diffraction (XRD) and Raman spectroscopy. XRD patterns were recorded using a
D8 Bruker diffractometer operating under the Bragg--Brentano geometry and
equipped with Cu anode providing X-rays with 1.5406~\AA\ wavelength. Raman
measurements were performed in a backscattering configuration at room
temperature, using an Ar$^{+}$ laser radiation ($\Lambda =$ 532~nm), focused
to a spot size around 0.8~$\mu $m on the sample surface. To avoid
laser-induced sample degradation, we performed data acquisition at 0.5\%
laser power for 300~s. Raman spectra were calibrated by a silicon reference
spectrum at room temperature.

To evaluate the quality of free-standing Ti$_{3}$C$_{2}$T$_{x}$ films and to
follow possible changes after thermal annealing performed during anelastic
measurements, we studied their structural properties, which are summarized
in Fig. S2. As-prepared free-standing Ti$_{3}$C$_{2}$T$_{x}$ film consists
of well delaminated flakes that are highly aligned along the $c$-direction,
and only peaks from $(00l)$ planes are visible in the XRD pattern (Figure
S2a).

\begin{figure}[t]
\begin{center}
\includegraphics[width=8.5cm]{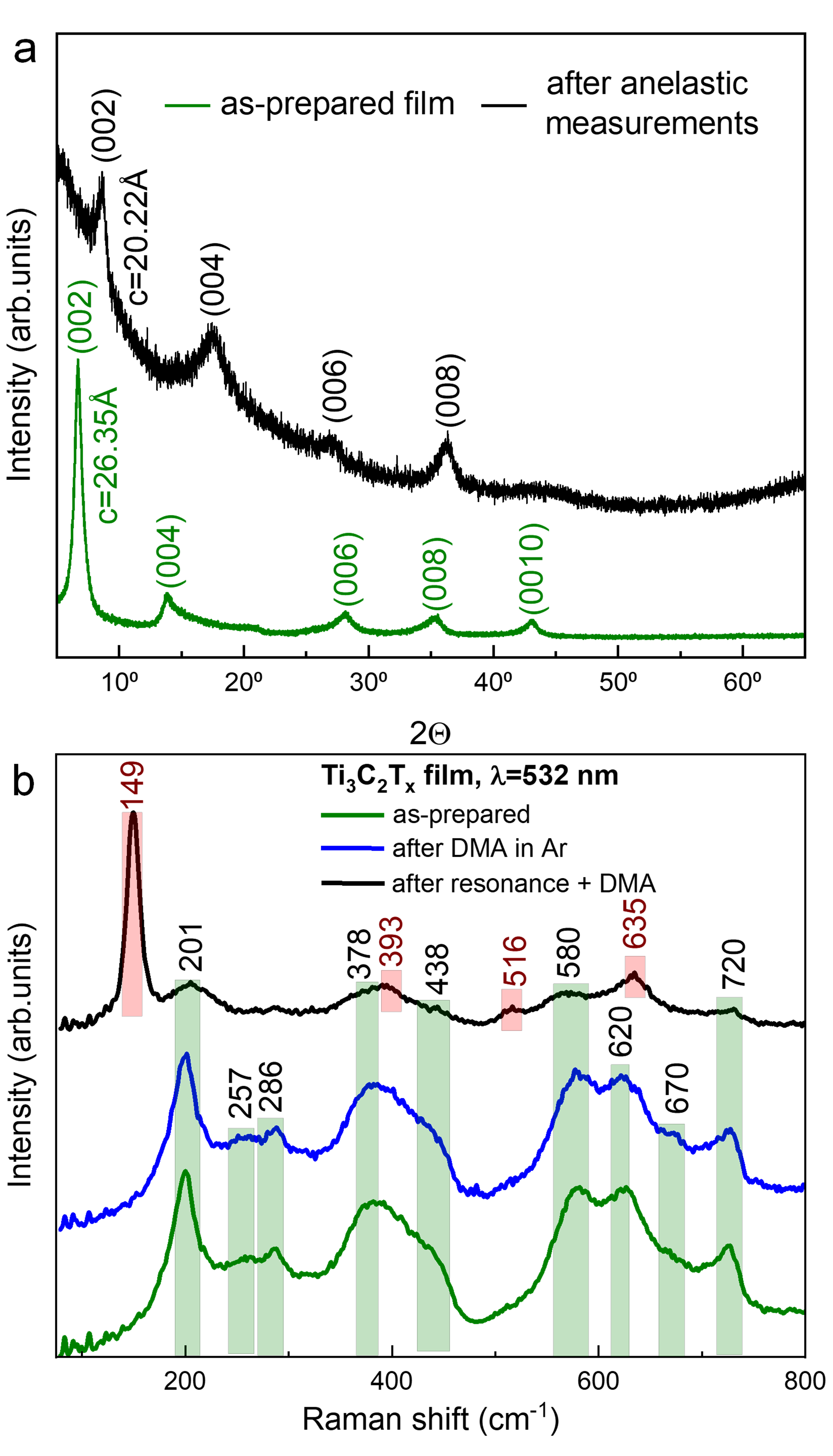}
\end{center}
\caption{XRD patterns (a) and
Raman spectra (b) of as-prepared Ti$_{3}$C$_{2}$T$_{x}$ film and after the
anelastic measurements. In the Raman spectra are marked the peaks
corresponding to Ti$_{3}$C$_{2}$T$_{x}$ in green, and the peaks
corresponding to TiO$_{2}$ (anatase phase) in pink.}
\label{fig-xrd-raman}
\end{figure}

The calculated $c$-lattice parameter of the as-prepared Ti$_{3}$C$_{2}$T$_{x}
$ film corresponds to 26.35 \AA . After the anneal at 700~${}^{\circ }$C and
the anelastic measurements in high vacuum, we observed a shift in the peaks
positions and a decrease in their intensity (Figure S2a). The shift in the
peaks position is due to the loss of intercalants and, as a consequence, a
decrease in the interplanar spacing between stacked MXene flakes. The $c$%
-lattice parameter of the Ti$_{3}$C$_{2}$T$_{x}$ film after the anelastic
measurements is 20.22~\AA , which is consistent with values reported for
annealed Ti$_{3}$C$_{2}$T$_{x}$ MXenes [\cite{ONO16}]. No new XRD peaks
appeared after the anelastic measurements, excluding a change in the phase
composition.

We then compare the Raman spectra of both films (Figure S2b). The Raman
spectrum of the as-prepared film shows peaks previously observed for Ti$_{3}$%
C$_{2}$T$_{x}$ MXenes [\cite{SG20}]. The main A$_{1g}$ mode, corresponding
to out-of-plane vibrations of titanium atoms in the outer layer, carbon and
surface groups, appears at 201~cm$^{-1}$. The second A$_{1g}$ mode,
corresponding to out-of-plane vibrations of carbon atoms, is located at
720~cm$^{-1}$. The broad peaks in the region of $230-470$~cm$^{-1}$ are
attributed to in-plane vibrations of surface functional groups bonded with
outer titanium atoms. The Raman spectrum measured on the film only subjected
to DMA in Ar (data not shown) looks the same as the spectrum measured on
as-prepared film. However, the sample from which the present results are
obtained, produced a different spectrum. We observed the appearance of new
peaks located at 149~cm$^{-1}$, 393~cm$^{-1}$, 516~cm$^{-1}$, and 635~cm$%
^{-1}$. These peaks correspond to the vibration modes of anatase. We note
that the E$_{1g}$ mode (149~cm$^{-1}$) of anatase has the highest intensity,
while the intensity of the peaks belonging to the Ti$_{3}$C$_{2}$T$_{x}$
structure decreased significantly after the cycles of measurements. Due to
the relatively shallow penetration depth of 532~nm light, we mainly probe 1~$%
\mu $m of the material. In this case, we attribute the formation of anatase
to surface oxidation of Ti$_{3}$C$_{2}$T$_{x}$, since no TiO$_{2}$ peaks
appear in XRD, probing the bulk of the film (Fig. S2a). We assume that
surface oxidation occurred when the film was heated in air saturated with
water.

\section{Elastic anomaly in the Landau theory}

A derivation is presented of the elastic anomaly at a ferroelectric
transition with spontaneous polarization $P$ and a polarization $p$ due to a
temperature independent intrinsic piezoelectric effect of the
non-centrosymmetric lattice under stress. The results in the absence of the
piezoelectric polarization $p$ are obtained setting $d=h=0$.

Interlayer polarization of the sliding FE, mainly perpendicular to the
layers: $P$

Intralyer piezoelectric polarization: $p$.

The Landau expansion of the Gibb's free energy $G=F-\sigma \varepsilon $ up
to the 6th power of $P$ is

\begin{eqnarray*}
G &=&\frac{a\left( T-T_{\mathrm{C}}\right) }{2}P^{2}+\frac{B}{4}P^{4}+\frac{C%
}{6}P^{6}-\frac{S^{0}}{2}\sigma ^{2}-Q\sigma P^{2}+\overset{G_{p}}{%
\overbrace{\frac{1}{2\chi _{p}}p^{2}-d\sigma p}}= \\
&=&\frac{1}{2}\underset{A^{\prime }}{\underbrace{\left[ a\left( T-T_{\mathrm{%
C}}\right) -2Q\sigma \right] }}P^{2}+\frac{B}{4}P^{4}+\frac{C}{6}P^{6}-\frac{%
S^{0}}{2}\sigma ^{2}+G_{p}
\end{eqnarray*}

Find the equilibrium value $p_{0}$ of $p$ in terms of the other variables,
in order to eliminate it from $G$ .

\begin{eqnarray*}
0 &=&\frac{\partial G}{\partial p}=p/\chi _{p}-d\sigma \\
p_{0} &=&\chi _{p}d\sigma
\end{eqnarray*}

Substituting into the Gibb's free energy

\begin{equation}
G_{p0}=G_{p}\left( P,p_{0},\sigma \right) =-\frac{\chi _{p}}{2}\left(
d\sigma \right) ^{2}  \label{Gp0}
\end{equation}

The spontaneous polarization $P_{0}$ in the stress-free state is similarly
found

\begin{equation}
0=A_{P}=\frac{\partial G_{0}}{\partial P}=\left( a\left( T-T_{\mathrm{C}%
}\right) -2Q\sigma \right) P+BP^{3}+CP^{5}
\end{equation}

\begin{equation}
0=A_{P}\left( \sigma =0\right) =P\left[ a\left( T-T_{\mathrm{C}}\right)
+BP^{2}+CP^{4}\right]  \label{Ap0=0}
\end{equation}

whose solution is
\begin{eqnarray}
P_{0}^{2} &=&\frac{B}{2C}\left( \pm \sqrt{1+\frac{4C}{B^{2}}a\left( T_{%
\mathrm{C}}-T\right) }-1\right) = \\
&=&\frac{B}{2C}\left( \pm \sqrt{1+\frac{T_{\mathrm{C}}-T}{\Delta T}}%
-1\right) ~%
\begin{array}{c}
+ \\
- \\
+%
\end{array}%
~%
\begin{array}{c}
B>0,~T<T_{\mathrm{C}} \\
B<0,~T<T_{\mathrm{C}}+\Delta T \\
B<0,~T_{\mathrm{C}}<T<T_{\mathrm{C}}+\Delta T%
\end{array}
\label{P0} \\
\Delta T &=&\frac{B^{2}}{4aC}
\end{eqnarray}

The sign must give $P_{0}^{2}>0$ at low $T:$

when $B>0$ it is $+$ with $\sqrt{}>1\rightarrow $ $T<T_{\mathrm{C}}$; the FE
transition occurs at $T_{\mathrm{C}}$;

when $B<0$ it is $-$ with $\sqrt{}>0\rightarrow $ $\frac{T_{\mathrm{C}}-T}{%
\Delta T}>-1\rightarrow $ $T<T_{\mathrm{C}}+\Delta T$;

when $B<0$ it is $+$ for $\sqrt{}-1<0\rightarrow $ $T_{\mathrm{C}%
}-T<0\rightarrow $ $T>T_{\mathrm{C}}$.

When $B<0$ there is possible coexistence of phases between $T_{\mathrm{C}}$
and $T_{\mathrm{C}}+\Delta T$.

The compliance is

\begin{equation*}
S=\frac{d\varepsilon }{d\sigma }=\frac{d}{d\sigma }\left( -\frac{\partial G}{%
\partial \sigma }\right) =\frac{d}{d\sigma }\left( QP^{2}+S^{0}\sigma +\chi
_{p}d^{2}\sigma \right) =S^{0}+\chi _{p}d^{2}+2QP\frac{dP}{d\sigma }
\end{equation*}

To find $\frac{\partial P_{0}}{\partial \sigma }$ differentiate $A_{P}\left(
\sigma \right) =0$ (Eq. (3.2), $P\equiv P_{0}$):

\begin{eqnarray*}
0=\frac{dA_{P}}{d\sigma }= &&\frac{d}{d\sigma }\left\{ P\left[ a\left( T-T_{%
\mathrm{C}}\right) +BP^{2}+CP^{4}-2Q\sigma \right] \right\} = \\
&=&\frac{\partial P_{0}}{\partial \sigma }\left\{ \overset{=0\text{ (Eq. 3.3
)}}{\overbrace{\left[ a\left( T-T_{\mathrm{C}}\right) +BP^{2}+CP^{4}\right] }%
}-\overset{=0\text{ stress-free state}}{\overbrace{2Q\sigma }}\right\} +P%
\left[ +2BP\frac{\partial P_{0}}{\partial \sigma }+4CP^{3}\frac{\partial
P_{0}}{\partial \sigma }-2Q\right] = \\
&=&P\frac{\partial P_{0}}{\partial \sigma }\left[ 2BP+4CP^{3}\right] -2QP=0
\\
\frac{\partial P_{0}}{\partial \sigma } &=&\frac{2QP}{2BP^{2}+4CP^{4}}
\end{eqnarray*}

When $P=P_{0}>0$

\begin{equation}
\Delta S=S-S^{0}=\chi _{p}d^{2}+\frac{2Q^{2}}{B+2CP_{0}^{2}}=\chi _{p}d^{2}+%
\frac{2Q^{2}}{\left\Vert B\right\Vert \sqrt{1+\frac{T_{\mathrm{C}}-T}{\Delta
T}}}
\end{equation}